\documentclass[aps,preprint]{revtex4}
\usepackage{amsmath}
\usepackage{blindtext}
\usepackage{extarrows}
\usepackage{braket}
\usepackage{pdfpages}
\usepackage{graphicx}
\usepackage{caption}
\usepackage{float}
\graphicspath{ {images/} }
\usepackage{amsmath,graphicx,color,xcolor,epsfig}
\usepackage{epstopdf}
\usepackage{hyperref}

\providecommand{\U}[1]{\protect\rule{.1in}{.1in}}

\begin{document}

\preprint{}
\title{Effects of intrinsic decoherence in multipartite system subjected to Kerr effect and parametric amplification with quantum correlations and estimation}
\author{M. Ibrahim$^{1}$}
\email{ibrahim@phys.qau.edu.pk}
\author{S. Jamal Anwar$^{2}$}
\author{S. Abdel-Khalek$^{3}$}
\author{M. Ramzan$^{2}$}
\author{M. Khalid Khan$^{2}$}
\affiliation{$^{1}$Department of Physics, University of Gujrat, Hafiz Hayat campus 50700, Pakistan}
\affiliation{$^{2}$Department of Physics, Quaid-i-Azam University, Islamabad}
\affiliation{$^{3}$Department of Mathematics, College of Science, Taif University, P.O. Box 11099, Taif 21944, Saudi Arabia}
\date{\today }

\begin{abstract}
We investigate the interplay between quantum correlations, quantified by the global quantum discord (GQD) and quantum Fisher information (QFI) in a multi- two-level system interacting with a single mode Fock field. Our model incorporates Kerr-like non-linearity effects, parametric amplification and intrinsic decoherence. We vary the cutoff photons in the system, system's dimensionality and varying amplification strengths under different magnitudes of Kerr effects, and analyze how these factors influence both correlations dynamics and parameter estimation. We observe a threshold-like behavior in photons number for transitioning from decoherence-dominated to coherence-enhanced regimes. These results intricate a balanced achieved between amplification strengths and Kerr effects and differing robustness properties of the GQD and QFI. These findings provide insights for optimizing non-linear atom-field systems for quantum information and metrology applications.\\

\end{abstract}

\maketitle

\section{\qquad Introduction}
Quantum entanglement and other quantum correlations are examples of fundamental phenomena that have been discovered via the study of quantum mechanics and have no classical counterparts. A crucial tool for many quantum information tasks, such as teleportation, quantum computation, and cryptography, is entanglement, which is a non-local correlation between two or more quantum systems \cite{ref1,ref2,ref3,ref4}. A significant barrier to creating reliable quantum devices is entanglement's vulnerability to external noise due to its fragility. Thus, one of the main challenges in the discipline is to comprehend and mitigate decoherence \cite{ref5,ref6,ref38}. Studies on noisy teleportation and open-system effects also highlight the detrimental impact of environmental interactions \cite{ref35}.  

Other types of quantum correlations exist and are often more resilient to decoherence. For instance, the global quantum discord (GQD) is a powerful measure of non-classical correlations in multipartite systems \cite{ref7}; unlike entanglement, which can vanish under particular conditions, the GQD can persist and may even be a more reliable resource for certain quantum tasks \cite{ref8,ref9,ref10,ref30}. Quantifying these correlations is crucial for characterizing the performance of quantum devices, and the theory of quantum metrology offers a framework for using quantum resources to achieve enhanced measurement precision \cite{ref11}. A key quantity in this field, the quantum Fisher information (QFI), serves as a theoretical bound on the precision of parameter estimation and can also act as an entanglement witness in multipartite systems \cite{ref12,ref13,ref14,ref15,ref31,ref33,ref36}. Recent studies further extended QFI analysis to noisy environments and generalized entanglement criteria \cite{ref37}.  

Nonlinear optical systems, such as those with the Kerr effect and parametric amplification (PA), are crucial platforms for generating and manipulating quantum states of light \cite{ref16,ref17,ref18,ref41}. Squeezed states and quantum Schrödinger cat states can be created via the Kerr effect, which characterizes the intensity-dependent change in a medium's refractive index \cite{ref19,kerr1, kerr2, kerr3}. Conversely, parametric amplifiers are frequently employed to amplify weak quantum signals with minimal additional noise and optical parametric amplification \cite{ref21,ref32}. Hybrid systems that couple optical and mechanical modes with Josephson parametric amplifiers provide additional means to boost entanglement and quantum coherence \cite{ref23,ref29}.  

While most studies on decoherence focus on the effects of an external environment (extrinsic decoherence), it is also important to consider models of intrinsic decoherence (ID) \cite{ref25}. Decoherence is a feature of the system's time development in this model that is independent of its interaction with an external bath \cite{ref26,ref27}. The Milburn model, which adds a stochastic element to the system's evolution at a fundamental level, explains its "intrinsic" nature \cite{ref28}. Experimental tests of intrinsic decoherence have also been reported \cite{ref44}. One important theoretical issue is to understand the impact of such a fundamental process on quantum states.  

The impact of intrinsic decoherence, parametric amplification, and the Kerr medium on quantum correlations and metrology have all been examined independently in earlier research. For example, their role in multipartite entanglement, quantum teleportation, and parameter estimation has been studied in various settings \cite{ref35,ref37,ref3}. Nevertheless, there is currently no thorough examination that looks into how these four elements, a multipartite system, a parametric amplifier, a Kerr medium, and intrinsic decoherence, combine to affect both the global quantum discord and quantum Fisher information. By offering a thorough theoretical examination of the dynamics of the GQD and QFI in such a complicated system, our work seeks to fill this gap. We will examine how the system's quantum correlations and ability to estimate parameters with high precision are affected by the interaction of these nonlinear effects and intrinsic decoherence. The structure of this paper is as follows: Section II introduces the model of atom–field interaction in the presence of a Kerr medium, parametric amplification and intrinsic decoherence. Section III focuses on multipartite quantum correlations and Fisher information, examining their significance within the framework of our system. Section IV presents the numerical results along with a thorough discussion of the observations. Finally, Section V concludes the study by summarizing the main findings and outlining possible directions for future research.

\section{Hamiltonian Model}

In this study, we explore an extended formulation of the Tavis-Cummings model \cite{tavis, 1-A}, a fundamental approach for examining multipartite quantum systems. The conventional version of this model describes two identical two-level atoms, denoted as A and B, coupled to a single-mode quantized electromagnetic field. Our generalization incorporates Kerr-type nonlinearity and degenerate parametric amplification, and we examine scenarios involving two, three, and four atoms confined within a cavity.

The Kerr effect originates from the third-order nonlinear optical response of a medium, producing an intensity-dependent phase shift in the field mode. This can be described by the refractive index relation
\begin{equation}
	n = n_0 + n_2 E^2,
\end{equation}
where $n_0$ is the linear refractive index, and $n_2$ represents the Kerr coefficient quantifying the magnitude of the nonlinear contribution.

Alongside the Kerr interaction, our model includes degenerate parametric amplification, a second-order nonlinear process in which a pump photon at frequency $2\omega$ splits into two photons of frequency $\omega$. This effect relies on a non-zero second-order susceptibility $\chi^{(2)}$ and is realized by driving a nonlinear medium with a classical pump field. In the Schrödinger picture, and setting $\hbar=1$, the Hamiltonian describing this process takes the form \cite{ref19}
\begin{equation}
	H_{PA} = -i \frac{\kappa}{2} \left(a^2 e^{2i\omega t} - a^{\dagger 2} e^{-2i\omega t} \right),
\end{equation}
where $a$ and $a^\dagger$ are, respectively, the annihilation and creation operators for the field, and $\kappa$ measures the pumping strength, proportional to both the pump amplitude and the nonlinear susceptibility.

Under the rotating wave approximation (RWA), rapidly oscillating terms are neglected, and the Hamiltonian in the interaction picture becomes
\begin{equation}
	H_{PA} = -i \frac{\kappa}{2} \left(a^2 - a^{\dagger 2} \right).
\end{equation}

Taking into account all relevant contributions, the complete Hamiltonian for $N$ two-level atoms interacting with the field is expressed as
\begin{equation}
	\hat{H}_T = \frac{\omega_0}{2} \sum_{i=1}^{N} \hat{\sigma}_i^z + \omega \hat{a}^\dagger \hat{a} + g \sum_{i=1}^{N} \left( \hat{a} \hat{\sigma}_i^+ + \hat{a}^\dagger \hat{\sigma}_i^- \right) + \chi \left( \hat{a}^\dagger \hat{a} \right)^2 - i \frac{\kappa}{2} \left( a^2 - a^{\dagger 2} \right),
\end{equation}
where $\omega_0$ and $\omega$ are, respectively, the atomic transition and field mode frequencies, $\hat{\sigma}_i^z$ and $\hat{\sigma}_i^\pm$ are the Pauli inversion and ladder operators for the $i$-th atom, $g$ denotes the coupling between atoms and field, and the $\chi$ term accounts for the Kerr nonlinearity.

The initial condition is chosen as a direct product of a partially mixed atomic state and a coherent state of the field:
\begin{equation}
	\hat{\rho}(0) = \left[ (1-p)|\psi\rangle\langle\psi| + p |g_1 g_2 \ldots g_N\rangle\langle g_1 g_2 \ldots g_N| \right] \otimes \hat{\rho}_E,
\end{equation}
where $0 \leq p \leq 1$ measures the mixedness. The pure atomic state $|\psi\rangle$ is
\begin{equation}
	|\psi\rangle = \cos(\theta) |g_1 g_2 \ldots g_N\rangle + \sin(\theta) |e_1 e_2 \ldots e_N\rangle,
\end{equation}
with $0 \leq \theta \leq \pi$, and $\hat{\rho}_E$ is the initial field state in the Fock basis:
\begin{equation}
	\hat{\rho}_E = \sum_{n} |n\rangle\langle n|.
\end{equation}

The composite system’s basis is
\begin{equation}
	\left\{ |\psi_i\rangle \right\} = \left\{ |s_1 s_2 \ldots s_N, n_c\rangle \mid s_j \in \{g,e\} \right\},
\end{equation}
where $n_c$ is the photon cutoff. The time-dependent state of the system is
\begin{equation}
	\hat{\rho}_{AF}(t)=\sum_{i,j}^{N}|\psi _{i}\rangle \langle \psi _{i}|\hat{\rho}(t)|\psi _{j}\rangle \langle \psi _{j}|,
\end{equation}
and within the Markovian framework \cite{ref25}, the evolution follows
\begin{equation}
	\dot{\hat{\rho}}(t)=-i[\hat{H},\hat{\rho}(t)]-\frac{\gamma}{2} [\hat{H},[\hat{H},\hat{\rho}(t)]],
	\label{equation 2r}
\end{equation}
where $\gamma$ is the intrinsic decoherence parameter. In the limit $\gamma \rightarrow 0$, Eq. (\ref{equation 2r}) reduces to the von Neumann equation. The general solution is
\begin{equation}
	\hat{\rho}(t)=\sum_{k=0}^{\infty}\frac{(\gamma t)^{k}}{k!}\hat{M}^{k}(t)\hat{\rho}(0)\hat{M}^{k\dagger}(t),
	\label{equation 3r}
\end{equation}
with
\begin{equation}
	\hat{M}^k(t)=\hat{H}^k\exp(-i\hat{H}t)\exp(-\gamma t \hat{H}^2 /2).
	\label{e4r}
\end{equation}
For $\gamma \neq 0$, the final state can be expressed in terms of the system’s eigenvalues as
\begin{equation}
	\hat{\rho}_{AF}(t)=\sum_{i,j;i\neq j}^{N}\exp\left[-\frac{\gamma t}{2}(E_i-E_j)^2-i(E_i-E_j)t\right] \bra{\psi_{i}}\hat{\rho}(0)\ket{\psi_{j}}\ket{\psi_i}\bra{\psi_j},
\end{equation}
where $E_i$, $E_j$ and $|\psi_i\rangle$, $|\psi_j\rangle$ are the corresponding eigenvalues and eigenvectors.
\section{Multipartite Quantum Correlations and Quantum Fisher Information}
Historically, the focus of quantum information theory was primarily on studying entanglement in bipartite systems. In the case of a composite system composed of two subsystems, $A$ and $B$, the quantity known as quantum discord, $D^{A \rightarrow B}$, measures non-classical correlations that persist even when no entanglement is present. It is defined as the difference between the quantum mutual information $I(\rho)$ and the classical part of the correlations $J(\rho)$, where the latter is minimized over the complete set of orthogonal projective measurements $\{\hat{\varPi}\}$ performed on subsystem $B$:  
\begin{equation}
	D^{A\rightarrow B}(\rho_{AB}) = \min_{\{\hat{\varPi}_B^j\}} \left[ I(\rho_{AB}) - J(\rho_{AB})_{\{\hat{\varPi}_B^j\}} \right].
\end{equation}

This concept has been broadened to address multipartite systems through the notion of the GQD. For an $N$-partite state, the GQD is expressed as:
\begin{equation}
	GQD(\rho_T) = \min_{\{\hat{\varPi}^j\}} \left[ S(\rho_T \| \hat{\varPi}(\rho_T)) - \sum_{j=1}^N S(\rho_j \| \hat{\varPi}_j(\rho_j)) \right],
\end{equation}
where $\rho_T$ denotes the global state of the system, $\rho_j$ is the reduced state of the $j$-th subsystem, and $S(\rho_1 \| \rho_2)$ is the relative entropy between two quantum states. This formulation provides a unified measure of quantum correlations across all subsystems.

For computational purposes, a more practical form of the GQD has been introduced \cite{camp}:
\begin{equation}
	GQD(\rho_T) = \min_{\{\varPi^k\}} \left\{ \sum_{j=1}^{N} \sum_{l=0}^{1} \tilde{\rho}_j^{ll} \log_2 \tilde{\rho}_j^{ll} - \sum_{k=0}^{2^N - 1} \tilde{\rho}_T^{kk} \log_2 \tilde{\rho}_T^{kk} \right\} + \sum_{j=1}^N S(\rho_j) - S(\rho_T),
\end{equation}
where $\tilde{\rho}_T^{kk} = \langle k | \hat{R}^\dagger \rho_T \hat{R} | k \rangle$ and $\tilde{\rho}_j^{ll} = \langle l | \hat{R}^\dagger \rho_j \hat{R} | l \rangle$. Here, $\hat{\varPi}^k = \hat{R} |k\rangle \langle k| \hat{R}^\dagger$ are the projectors, and the rotation operator $\hat{R}$ is defined as $\hat{R} = \bigotimes_{j=1}^N \hat{R}_j(\theta_j, \phi_j)$ with $\hat{R}_j(\theta_j, \phi_j) = \cos\theta_j \hat{1} + i \sin\theta_j \cos\phi_j \hat{\sigma}_y + i \sin\theta_j \sin\phi_j \hat{\sigma}_x$.

In the field of quantum metrology, the QFI serves as a central tool for estimating an unknown parameter $\theta$ with optimal precision. Its classical counterpart, the Fisher information (CFI), is defined as:
\begin{equation}
	I_\Phi = \sum_i p_i(\theta) \left( \frac{\partial}{\partial\theta} \ln p_i(\theta) \right)^2,
\end{equation}
where $p_i(\theta)$ is the probability of obtaining the $i$-th measurement result, dependent on $\theta$.

The QFI generalizes this idea to the quantum regime, setting the ultimate bound on estimation precision. It is given by:
\begin{equation}
	F_\Phi = \text{Tr}[\rho(\theta) D^2],
\end{equation}
where $D$ is the symmetric logarithmic derivative (SLD), determined through:
\begin{equation}
	\frac{d\rho(\theta)}{d\theta} = \frac{1}{2} \left[ \rho(\theta) D + D \rho(\theta) \right].
\end{equation}

If $\rho_\theta$ is diagonalized as
\begin{equation}
	\rho_\theta = \sum_K \lambda_K |k\rangle \langle k|,
\end{equation}
then the QFI can be written as:
\begin{equation}
	F_\theta = \sum_k \frac{(\partial_\theta \lambda_k)^2}{\lambda_k} + 2 \sum_{k,k'} \frac{(\lambda_k - \lambda_{k'})^2}{\lambda_k + \lambda_{k'}} \left| \langle k | \partial_\theta k' \rangle \right|^2,
\end{equation}
with the constraints $\lambda_k > 0$ and $\lambda_k + \lambda_{k'} > 0$. The first summation corresponds to the classical contribution, while the second captures the quantum part.

To determine the average quantum Fisher information (AQFI) in a composite setup, a partial trace over the field degrees of freedom is taken. For a bipartite state $\rho_{AB}$, AQFI is expressed as:
\begin{equation}
	I_{QF}(t) = \text{Tr}[\rho_{AB}(\theta, t) D(\theta, t)^2],
\end{equation}
where the SLD $D(\theta, t)$ obeys:
\begin{equation}
	\frac{\partial \rho_{AB}(\theta, t)}{\partial \theta} = \frac{1}{2} \left[ D(\theta, t) \rho_{AB}(\theta, t) + \rho_{AB}(\theta, t) D(\theta, t) \right].
\end{equation}

\section{Numerical \textbf{results and discussions}}
We numerically solve the model composed of up to four two-level atomic system interacting with the Fock field for the GQD and QFI. We assume that the system is subjected to intrinsic decoherence of value $\gamma=0.05$. We take the scaled time step size of 0.05.
\subsection{The Effects of cutoff photons on the quantum correlation and Fisher information}
\begin{figure}[H]
	\centering
	\includegraphics[width=5.5in]{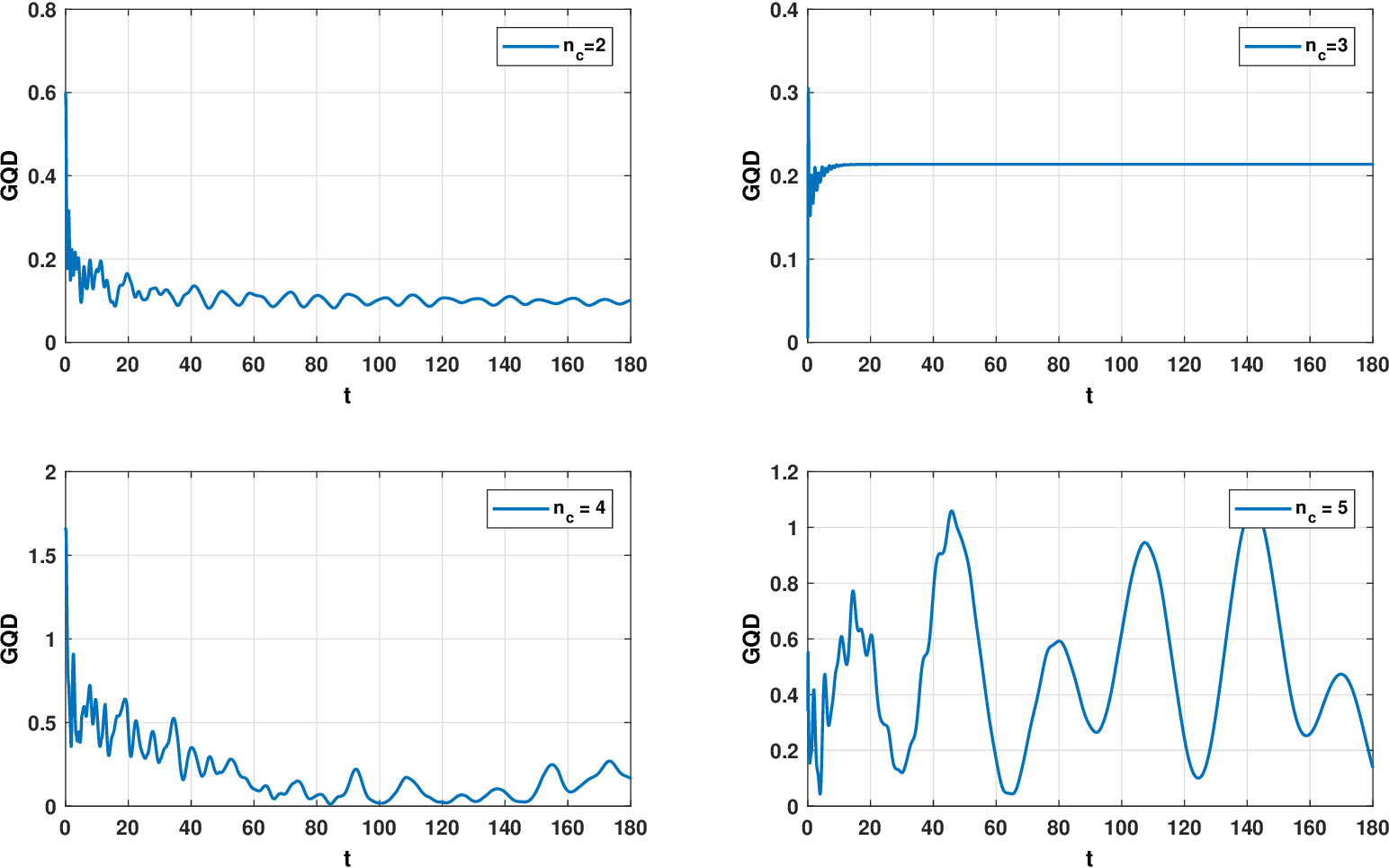}
	\caption{(color online) The interplay between the cutoff photons in the presence of intrinsic decoherence is shown. Higher cutoff photons assist the correlations dynamics in non-linear media. We have taken the parametric amplifier amplitude $\kappa=1$, the Kerr parameter $\chi=1$, and the value of intrinsic decoherence is $\gamma=0.05$. The initial state parameters chosen for this figure are $p=0.5$ and $\theta=\pi/4$. We have consider the case of two two-level atomic system.}
	\label{fig1}
\end{figure}

\begin{figure}[H]
	\centering
	\includegraphics[width=5.5in]{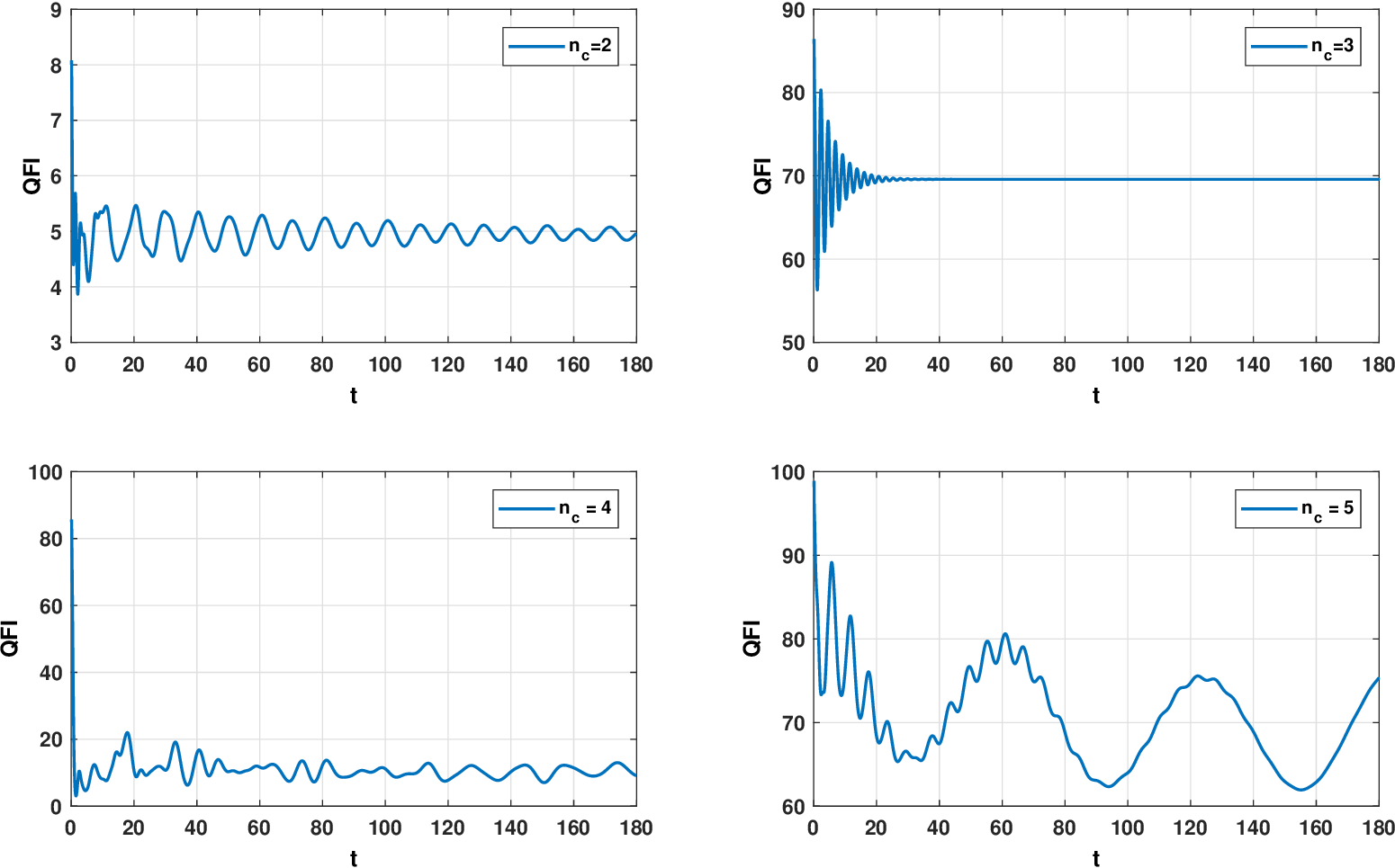}
	\caption{(color online) The dynamics of the QFI for different cutoff photons is shown in this figure. The system is evolved in the presence of intrinsic decoherence of value $\gamma=0.05$. The QFI is estimated around the initial state parameter $\theta$ at value $\pi/4$. We note that cutoff photons have strong impact on the parameter estimation. The other values are chosen as given in \ref{fig1}.}
	\label{fig2}
\end{figure}

In Fig. (\ref{fig1}), we study the temporal evolution of the GQD for a two two-level atomic system interacting with a single mode Fock field. We also consider that the system is under the collective influence of a Kerr like non-linearity and parametric amplification. The cutoff photons number are varied from 2 to 5 and we keep both the Kerr and parametric amplification amplitude fixed at 1. \\
For the case of $n_c =2$, the GQD starting value initiate at relatively high value, indicating that there are strong initial quantum correlations in the system. However, the effect of intrinsic decoherence quickly come into play. The GQD undergoes a rapid decay. The dynamics of the GQD at later time shows irregular and damped oscillations. These oscillations die out over time and the GQD tends to settle at a non-zero small value. Due to limited number of interacting photons, the resulting short lived coherence and suppressed revivals of quantum correlations. The resulting decay pattern of the GQD for the system confined to a small photons clarifies the dynamics. For the case of $n_c=3$, the GQD still starts from a moderate value and displays an initial fluctuations. It stabilizes more quickly compared to $n_c=2$ case. After a brief period of oscillatory behavior in the dynamics, the GQD achieve a steady state value with minimal oscillations. This steady state behavior in the GQD dynamics is due to the effect of a stable interplay between the intrinsic decoherence and slightly increased photons in the system. This steady state value of quantum correlations after brief oscillations for $n_c=3$ is slightly improved as compared to $n_c=2$ case. For the case of cutoff photons $n_c=4$, a significant change in the dynamics of the GQD is observed. There is a pronounced initial rise that reaches a peak value exceeding 1.5 is observed. This substantial enhancement in the GQD is followed by damped oscillations. The decay in the dynamics of the GQD is much slower that the cases of lower cutoff photon case. This indicates the accessibility of the system with the greater number of cutoff photons facilitate the revivals of the quantum correlations and the system maintains the GQD over larger scaled times. Increasing cutoff photons can mitigate the effects of intrinsic decoherence for some extent. When we increase the cutoff photons to $n_c=5$, the GQD dynamics for this cutoff value exhibits a qualitatively different behavior. Instead of decaying and then stabilizing, the GQD shows large-amplitude oscillations. The GQD displays an overall increasing trend, reaching a maximum around $t=140$ scaled time. This suggest that coherent dynamical nature of atom-field are now dominating over the decoherence effects. This nature is due to the aid by the higher cutoff photons. The enhanced interaction of the system with the field in the presence of non-linearity effects enable sustained growth of the GQD even in the presence of intrinsic decoherence. The parametric amplifier continuously pumps the energy into the system, while the Kerr interaction with the system facilitate entanglement across the subsystems. This case of cutoff photons demonstrates that beyond a certain photon number threshold, the system can exhibits robust and even growing quantum correlations. This also indicates that there exist a transition from a decoherence-dominated regime to a coherence-enhanced regime. 
Fig. (\ref{fig2}) shows the dynamics of the QFI for each cutoff photons cases studied for the GQD. The system is subjected to intrinsic decoherence of value $\gamma=0.05$. The QFI is analyzed around the initial state parameter $\theta$ around the value $\pi/4$. The Kerr and parametric amplification parameters both are set at 1.\\
For the case of $n_c=2$, the QFI value starts around $4$-$5$ and its dynamics quickly settles into small-amplitude oscillations. The oscillations oscillate about a mean value of approximately 5. These oscillations are nearly periodic and the amplitude of oscillations is relatively small. This shows that the system sensitivity to the initial state parameter estimation of $\theta$ remains modest in the dynamics. This is also due to relatively small cutoff photons interaction, which limits the effective atom-field correlations that can develop. Intrinsic decoherence suppresses the parameter sensitively almost immediately in the dynamics. As a result of decoherence, the system quickly enters a steady state oscillatory regime which has low but stable QFI. For the case of $n_c=3$, the QFI exhibits a large initial peak over $80$. This is followed by a fast oscillations that quickly damp to a stable value near $70$. This high steady state value indicates that even the decoherence removes the fast oscillations of the QFI, a significant amount of the parameter estimation of $\theta$ remains in the system. Allowing these cutoff photons increase the accessible field states to the system, enhances the QFI in the presence of the parametric amplification and Kerr non-linearity. For the case of $n_c=4$, the dynamics show a dramatic initial spike close to $100$, indicating high sensitivity to $\theta$ parameter in the early stage. However, this is followed by a rapid collapse to the low values between $5$ and $15$. The larger photon cutoff provides a richer interaction of the system with the field. This case of cutoff photons momentarily supports a strong multipartite correlations and sharp phase sensitivity. This large sensitivity is yet fragile and vulnerable to intrinsic decoherence. For the case of $n_c=5$, the QFI displays sustained large-amplitude oscillations between about $60$ and $90$ over the entire time range. The oscillations show slow and long-period revivals. Unlike lower cutoff cases, the high photon number works in creating strong and long lived QFI values that are less affected by decoherence. As a result of high cutoff photons, the system enters into a regime that facilitates the high sensitive to the estimation of parameter $\theta$, showing a threshold-like shift from decoherence-limited to a coherence-enhanced quantum metrological behavior.  
\subsection{Effects of system's Hilbert space on the dynamics of the quantifiers}
\begin{figure}[H]
	\centering
	\includegraphics[width=4.5in]{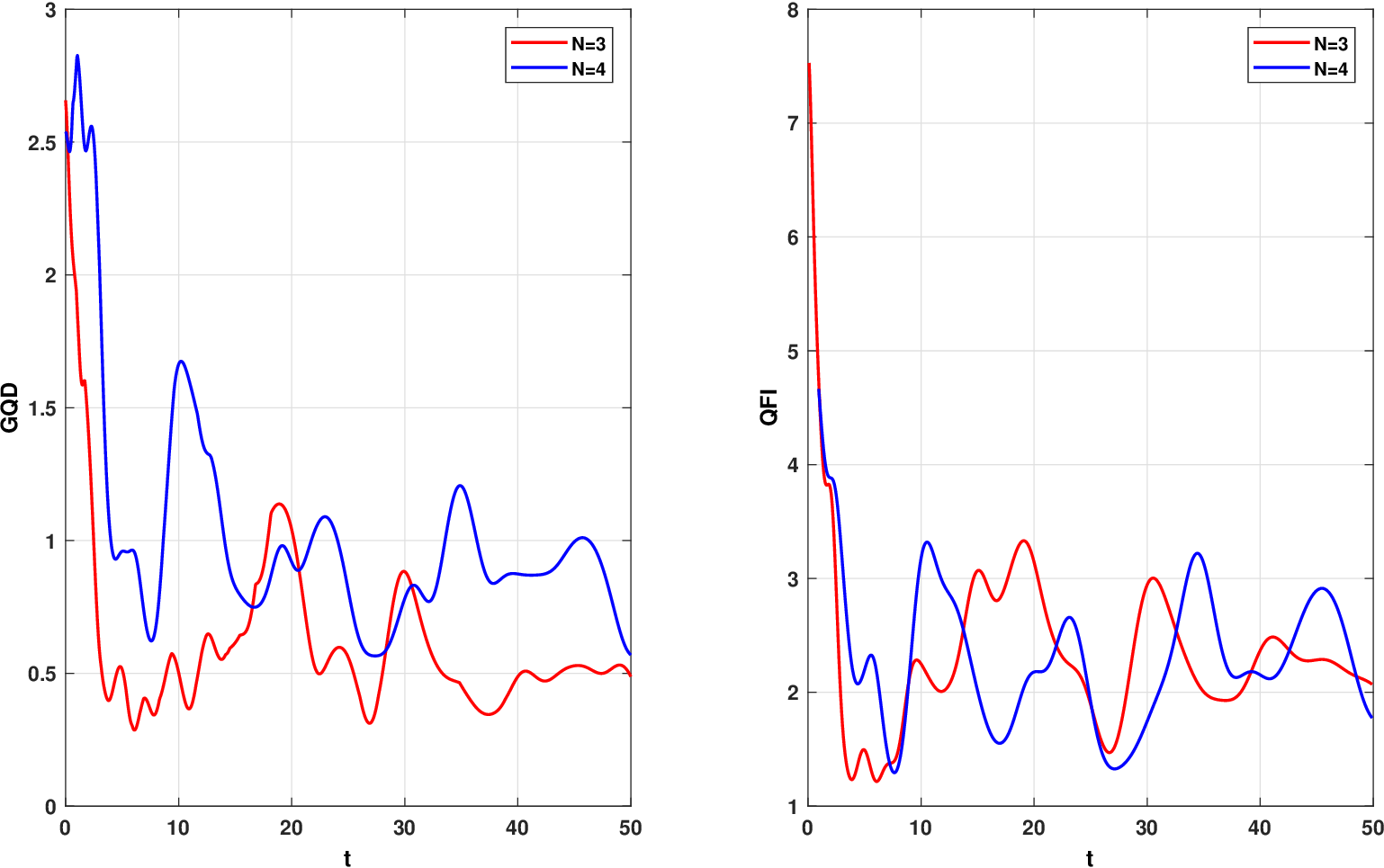}
	\caption{(color online) The effect of system size on the quantum correlations and parameter estimation in the presence of Kerr nonlinear medium and parametric amplification is displayed in this figure. The quantum correlations are enhanced while parameter estimation are not improved while the system's Hilbert space is increased. For this Figure we have taken $n_c=2$ and other parameters are same as in Figs. (\ref{fig1})-(\ref{fig2}). The QFI is estimated around $\theta=\pi/4$}.
	\label{fig3}
\end{figure}

In this section, we analyze the GQD and QFI for the system of $N=3$ and $N=4$ two-level atomic systems. The system is interacting with a single mode field under fixed cutoff photons $n_c=2$. We set Kerr and parametric amplification factors to unity. The intrinsic decoherence rate is taken to be $\gamma=0.05$. In this section, the QFI is estimated at $\theta=\pi/4$.\\
The left panel of Fig. (\ref{fig3}) shows the GQD for $N=3$ and $N=4$. For both cases, the GQD start with high initial values, that rapidly decay due to the decoherence. However, $N=4$ case reaches a slightly higher initial peak ($\approx 2.6-2.8$) than $N=3$ ($\approx 2.4-2.6$). Furthermore, $N=4$ maintains larger oscillation amplitude throughout the evolution. To understand this behavior we say can say that adding an extra atomic system expands the multipartite Hilbert space, enabling stronger initial correlations in the dynamics through the combined effects of Kerr and parametric amplification mechanisms. Intrinsic decoherence damps the GQD oscillations, leading both systems to a regime of damped oscillations. For the case of $N=3$, these oscillations are more pronounced and the mean value of the oscillations as time progresses remains higher ($0.6-1.0$) compared to $N=3$ ($0.3-0.6$). This indicates that there is greater robustness of quantum correlations in the larger atomic ensambles. For the case of the QFI, both systems exhibit large QFI values with peak at approximately $7$. The QFI drops quickly for both systems under intrinsic decoherence. Initially, the behavior favors $N=4$ system in terms of oscillations amplitude, yet $N=3$ occasionally surpasses $N=4$ at certain times. This reflects that the QFI depends on how coherence is aligned with the parameter $\theta$ around $\pi/4$ estimation, and increasing $N$ does not grantee the enhancement of estimation in the dynamics. Over long time, both curves oscillate around $\approx 2-3$. Overall a larger ensamble of two-level system in the Kerr medium and parametric amplification does not grantee a steady-time advantage in the presence of intrinsic decoherence of the parameter estimation. 
\subsection{Investigation of parametric amplification pumping strength for variable medium's Kerr effects}
\begin{figure}[H]
	\centering
	\includegraphics[width=5.5in]{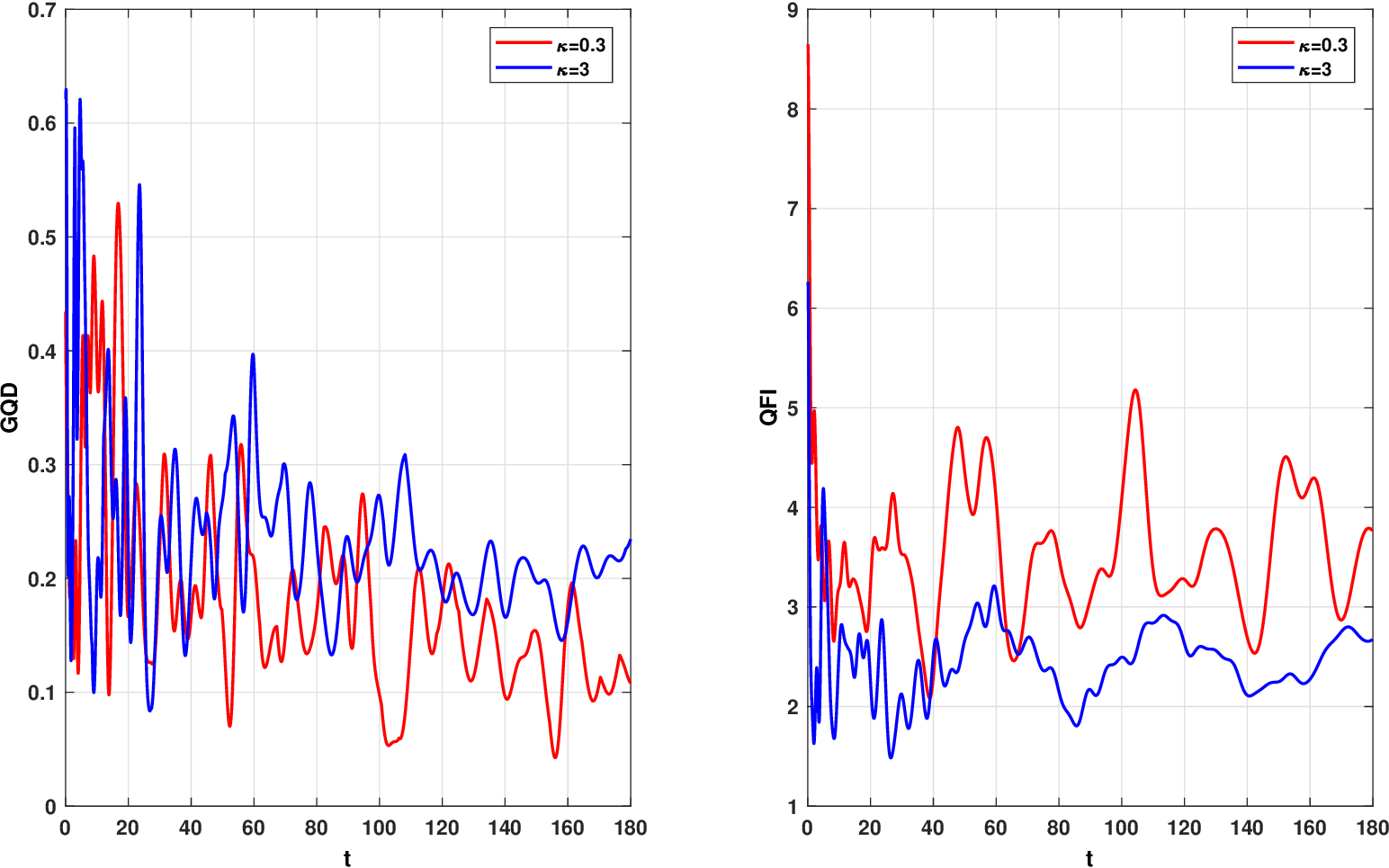}
	\caption{(color online) Figure shows the dynamics of the GQD and QFI for the case of Kerr and parametric amplification presence. The Kerr effect parameter is set to $\chi=0.3$. Intrinsic decoherence is also in effect with $\gamma=0.05$ and we take $N=2$. We note that the strong parametric amplification pump amplitude $\kappa$ assists the dynamics of the GQD. The initial state parameters values is taken same as in Figures above. The QFI is estimated around $\theta=\pi/4$.}
	\label{fig4}
\end{figure}

\begin{figure}[H]
	\centering
	\includegraphics[width=5.5in]{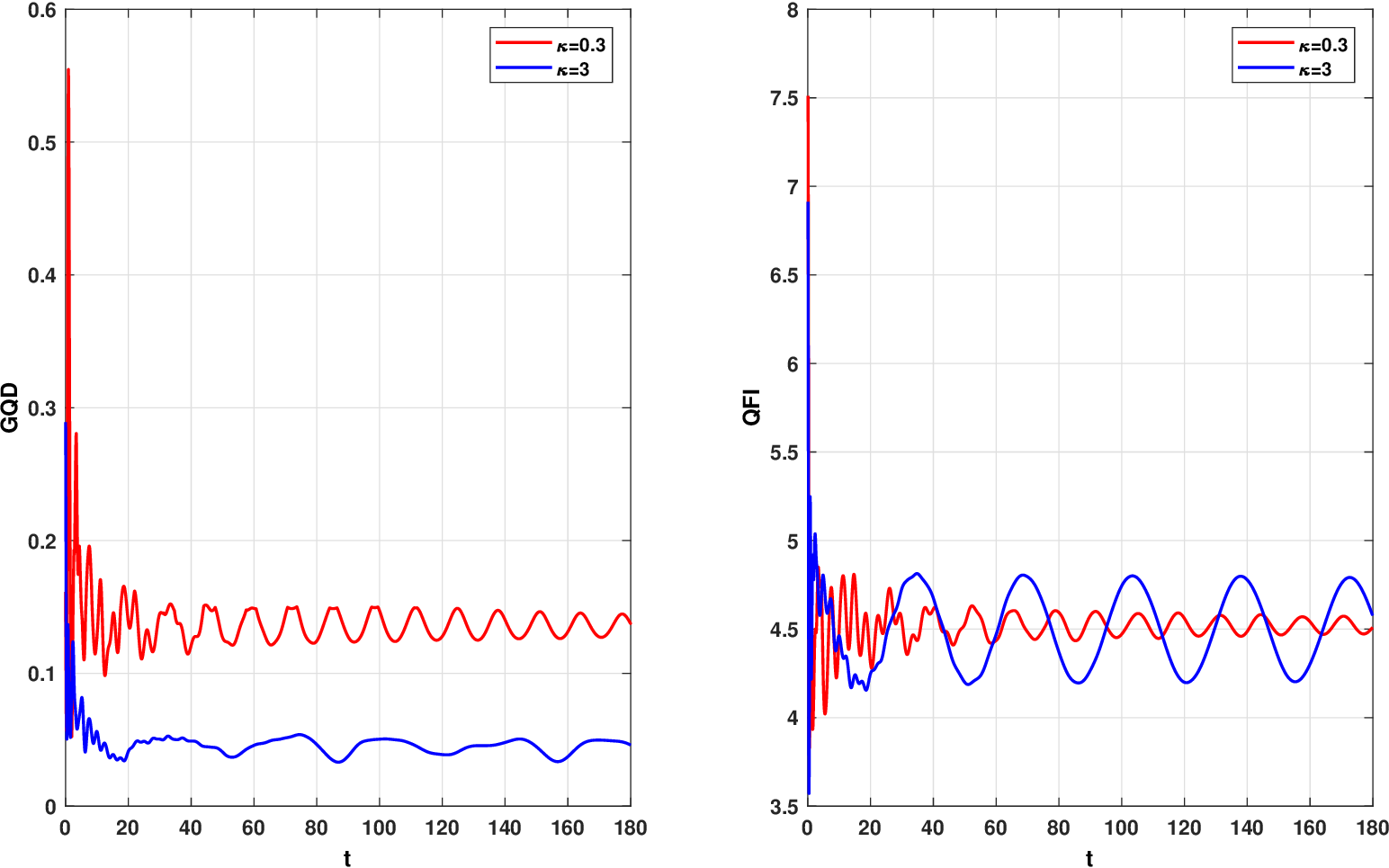}
	\caption{(color online) The figure shows the dynamics of the GQD and QFI for a system with N=2. The system is subjected to Kerr nonlinear effect with parametric amplification. We find that stronger Kerr effect with stronger parametric amplification amplitude favors metrological sensitivity of a initial state parameter. }
	\label{fig5}
\end{figure}

In this section we investigate the influence of the strength of parametric amplification pump amplitude parameter $\kappa$ on the dynamics of the GQD and QFI for the case of Kerr parameter value fixed at $\chi=0.3$. We study the case of $N=2$ and with intrinsic decoherence present at $\gamma=0.05$. The change in the quantum correlations and parameter estimation can be attributed primarily to the variations in the amplification strengths. The strength of parametric amplification acts as a source of pumping strength of photons in the cavity mode. \\
The left panel of Fig. (\ref{fig4}) shows the dynamics of the GQD for both $\kappa $ values. The system exhibits an initial oscillatory decay of the quantum correlations due to the presence of intrinsic decoherence. At early time ($t<20$), the smaller amplification ($\kappa=0.3$) shows slightly larger GQD amplitude. But as time progresses, the stronger amplification ($\kappa=3$) maintains a higher average value and more persistent oscillations. For a weak Kerr value ($\chi=0.3$), a strong $\kappa$ increases the photon injection strength which help replenish atom and field correlations even the decoherence acts to destroy them. This causes a noticeable long-time advantage for $\kappa=3$ where the quantum correlations remains around $0.2$ and $0.3$ compared to $0.1$ to $0.2$ for $\kappa=0.3$. The higher $\kappa$ is, therefore, acts as a stabilizer for multipartite correlations in the presence of noise. For the case of the QFI (right panel of (\ref{fig4})), its dynamics shows a different trend. For small time, both curves start with large peaks. But for $\kappa=0.3$, the values reach higher maxima especially with revival like events (e.g. at $t\approx 40, 100, 140$). This suggests that weak amplification factor ($\kappa=0.3$) can produce interference that assists the chosen parameter $\theta$ around estimation $\pi/4$. On the other hand, for strong amplification factor ($\kappa=3$), the QFI exhibits smaller revivals amplitudes stabilizing around $2$-$3$ as time progresses. The stronger parametric pump amplitude adds more noise-like excitation that negatively affects optimal parameter sensitivity. Therefore we note that strong $\kappa$ benefits long-time stabilizing of the GQD but does not necessarily enhances, even suppresses the QFI compared to weak amplifier pump. 
In Fig. (\ref{fig5}), we analyze the time evolution of the GQD and QFI for different strengths of parametric amplification factor, $\kappa=0.3$ and $\kappa=3$, in the presence of the non-linear Kerr medium of parameter $\chi=3$. The value of intrinsic decoherence is taken as $\gamma=0.05$. We estimate the QFI around $\theta=\pi/4$.\\
The left panel of Fig. (\ref{fig5}) shows the GQD dynamics. At $t=0$, the GQD starts at approximately $0.55$ for $\kappa=0.3$ and $0.3$ for $\kappa=3$. For both cases, the GQD undergoes a rapid decay in the first $t\approx 10$ scaled time due to the combined effects of intrinsic decoherence and Kerr induced dephasing. As time progresses, the GQD for $\kappa=0.3$ stabilizes around $0.15$-$0.18$ with small oscillations of amplitude $\approx 0.01$. On the other hand, for $\kappa=3$, the GQD settles to a significantly lower plateau of $0.045$-$0.06$. This shows that in the presence of a stronger Kerr effect, larger $\kappa $ values suppresses long-time correlations rather that sustaining. This behavior is in contrast to the relative weak Kerr parameter regime. The right side panel of Fig. (\ref{fig5}) shows the QFI dynamics. Initially the QFI reaches about $7.5$ for $\kappa=0.3$ and $6.9$ for $\kappa=3$. Both cases decay towards a steady state oscillatory regime, but the nature of the behavior differs. For $\kappa=0.3$, the QFI weakly fluctuates between $4.4$ and $4.6$. On the other hand, for $\kappa=3$, it exhibits pronounced oscillations that are periodic in nature. The amplitude of oscillations has value $\approx 0.6$, swinging between $4.0$ and $5.2$ with well defined period of $\approx 40$ scaled time. This suggests that strong parametric amplification in the presence of large Kerr effect enhances the coherent dynamics leading to more pronounced oscillatory revival amplitude of the QFI, even though the average value is comparable to the weaker $\kappa $ cases. The results indicates that for relative stronger Kerr effect, smaller $\kappa$ support higher long-time quantum correlations while larger $\kappa$ favors stronger periodic modulations in metrological sensitivity. 
\section{Conclusions}
In this work we have numerically analyzed the dynamics of the GQD and QFI for the system of up to four two-level atomic systems interacting with a single mode Fock field under the combined influence of Kerr non-linearity, parametric amplification, and intrinsic decoherence. We explored a wide variations in the model parameters including changing in cutoff photons, the dimensions of the system, and the interplay between weak and strong parametric amplifications strengths in both low and high Kerr effects. The results reveal rich dynamical features such as non-monotonic decay pattern, steady state behaviors, large-amplitude revivals, and parameter dependent enhancement or suppression of the quantum correlations and Fisher information.\\
Our finding showed that by increasing the cutoff photons in the system, can transitions the system from a decoherence-limited to a sustained coherence regime. This increase in cutoff photons enabled sustained or even growing quantum correlations in the presence of intrinsic decoherence. However, the QFI response to photon number and the atomic size was more subtle, often proving fragile against the decoherence. In the amplification-Kerr interplay, we found that strong parametric pumping enhances long-time GQD for weak Kerr, but can suppress it under strong Kerr effect, whereas the QFI benefited from sustained revival amplitude in high Kerr regime. These results emphasize that optimizing such systems requires a careful choice of photons numbers, amplification strengths, and Kerr effect behavior to achieve the desired balance between robust correlations and parameter estimation.\\
Looking forward, this work can be extended to several promising research directions. We can extend the analysis to finite temperature fields could provide a more realistic open-system environments. We can incorporate time dependent amplification schemes might enable active stabilization of quantum correlations against intrinsic decoherence. Furthermore, generalizing the model to multi-mode field or optomechanical systems could broaden the applicability of these findings to practical implementations in quantum communications, sensing and computations, where both quantum correlations and metrological precision are crucial.\\

\end{document}